\documentclass[useAMS,usenatbib]{mn2e}

\usepackage{times}
\usepackage{graphics,epsfig}
\usepackage{graphicx}
\usepackage{amsmath}
\usepackage{amssymb}

\title[The shapes of dwarf irregulars]{The intrinsic shapes of dwarf irregular galaxies}

\author[Roychowdhury et al.]{Sambit Roychowdhury,$^{1}$\thanks{E-mail: sambit@ncra.tifr.res.in (SR); chengalu@ncra.tifr.res.in (JNC); ikar@sao.ru (IDK); kei@sao.ru (EIK)} Jayaram N. Chengalur,$^{1\star}$ Igor D. Karachentsev,$^{2\star}$ 
\newauthor Elena I. Kaisina$^{2\star}$\\
       \\ 
       $^{1}$NCRA-TIFR, Post Bag 3, Ganeshkhind, Pune 411 007, India\\
       $^{2}$Special Astrophysical Observatory, Russian Academy of Sciences, N. Arkhyz, KChR 369167, Russia}

\begin{document}
\date{}

\pagerange{\pageref{firstpage}--\pageref{lastpage}} \pubyear{}

\maketitle

\label{firstpage}

\begin{abstract}
We use the measured B band axial ratios of galaxies from an updated catalog of Local Volume galaxies to determine the intrinsic shape of dwarf irregular galaxies (de Vacouleurs' morphological types 8, 9 and 10). We find that the shapes change systematically with luminosity, with fainter galaxies being thicker. In particular, we divide our sample into sub-samples and find that the most luminous dwarfs ($-19.6 < {\rm M_{\rm B}} < -14.8$) have thin discs (thickness $\sim 0.2$), with the disc being slightly elliptical (axial ratio $\sim 0.8$). At intermediate luminosity, viz. $-14.8 < {\rm M_{\rm B}} < -12.6$, the galaxies are still characterized by elliptical discs (axial ratio $\sim 0.7$), but the discs are somewhat thicker (thickness $\sim 0.4$). The faintest dwarfs, viz. those with $-12.6 < {\rm M_{\rm B}} < -6.7$ are well described as being oblate spheroids with an axial ratio $\sim 0.5$. The increasing thickness of the stellar discs of dwarf irregulars with decreasing luminosity is compatible with the increasing ratio of velocity dispersion to rotational velocity with decreasing galaxy size.
\end{abstract}


\section{Introduction}
\label{sec:optaxrat_int}

The intrinsic three dimensional shapes of galaxies provide constraints for both dynamical models of galaxies as well as models of their evolutionary history. Early simulations of hierarchical structure formation showed that baryons accrete and cool within galactic dark matter halos while acquiring some of the angular momentum of the dark matter via dynamical friction. As a result the baryons settle into cold thin tightly bound discs \citep{fal80,nav95}. More recent models including gas dissipation and feedback from star formation, as well the effect of mergers, result in the formation of more realistic structures for both disc and spheroidal galaxies \citep{tot92,nav97,wei98,DeL06,bou07,gov10,bro11,sal12}. For disc galaxies, knowledge of the shape, particularly the intrinsic thickness of galaxy discs is also required to determine the inclination from the observed axial ratio. The inclination angle in turn, is an important input for determining the surface brightness profiles, modeling the dark matter halo etc.

Although theoretical modeling of galaxy shapes is a relatively recent development, there has been a long history of observational studies to determine the shapes of galaxies as a function of their morphological type.  \citet{heidmann72} showed that as one goes from morphological type Sa galaxies to type Sd galaxies the stellar discs get thinner, but there is a sharp increase in thickness as on goes from Sd galaxies to Irregular galaxies (their Figures 1 and 2). \citet{bin81} also showed that intrinsic axial ratios of optical discs of Sdm, Sm and Im types \citep[types 8, 9 and 10 as defined in][]{deVac91} in the {\it Second Reference Catalogue} have a wide distribution between $\sim$0.2 and $\sim$0.8 (their Figure 7) compared to the thin optical discs of earlier type spirals. \citet{vandB88} showed that not only are the stellar distributions of irregular galaxies not well represented by spheroids with intrinsic axial ratios of $\sim$0.2 as spiral galaxies are, but also that dwarf irregular galaxies are triaxial systems with axial ratios of 1:0.85:0.4. \citet{lam92} showed that the bulges of bright elliptical galaxies were mildly triaxial structures, with two axes having comparable scale length, while the third is slightly smaller. Interestingly, they also found that the stellar discs of bright spirals were also triaxial (not merely thin oblate spheroids) with the two larger axes having slightly differing scale lengths.
In a recent study using the Sloan Digital Sky Survey data release 6 (SDSS-DR6), \citet{pad08} showed that though the intrinsic shapes of elliptical galaxies on the whole are consistent with oblate spheroids, when looked at separately low luminosity ellipticals are consistent with prolate spheroids and high luminosity ones with oblate spheroids. As for spiral galaxies, \citet{pad08} found that obscuration due to dust makes spiral discs appear rounder than they actually are. They determine that spirals have thin discs with slight face-on ellipticity, and the discs become thicker and more circular with increasing luminosity.

There have also been a number of studies of the intrinsic shape of dwarf galaxies.  \citet{sta92} analyzed data from 483 galaxies listed as `dwarf' or `dwarf irregular' in the UGC catalogue and found them to be thick oblate spheroids with mean intrinsic axial ratio of 0.57.
\citet{sun98} showed that faint dwarf irregulars and blue compact dwarfs have similar intrinsic shapes: highly elliptical thick discs.
In a recent study \citet{san10} assumed galaxies to be oblate spheroids and investigated the variation of the thickness with magnitude and stellar mass. They note that the thinnest discs are found around ${\rm M_*~\approx~2~\times~10^9~M_{\odot}}$~or equivalently around ${\rm M_i~\sim~-18}$, and the discs become progressively fatter on either side.  The brighter galaxies in their sample were drawn from SDSS data release 7 (SDSS-DR7), but galaxies with ${\rm M_B~>~-14.5}$ were taken from the catalogue of \citet{kar04}.

Here we focus on the intrinsic shape of the stellar distribution in dwarf irregular galaxies. These galaxies are the most numerous among all star forming galaxies \citep[e.g.][]{gal84}. We use data from an updated survey of the Local Volume galaxies, which makes the sample for this study the largest among studies of the intrinsic shapes of dwarf galaxies, for galaxies with ${\rm M_B~\ge~-14.5}$. We do not assume that the intrinsic shapes are oblate spheroids, but try and determine the shape from the data itself. 

\section{Observed axial ratio distributions}
\label{sec:optaxrat_samp}

\begin{figure}
\begin{center}
\psfig{file=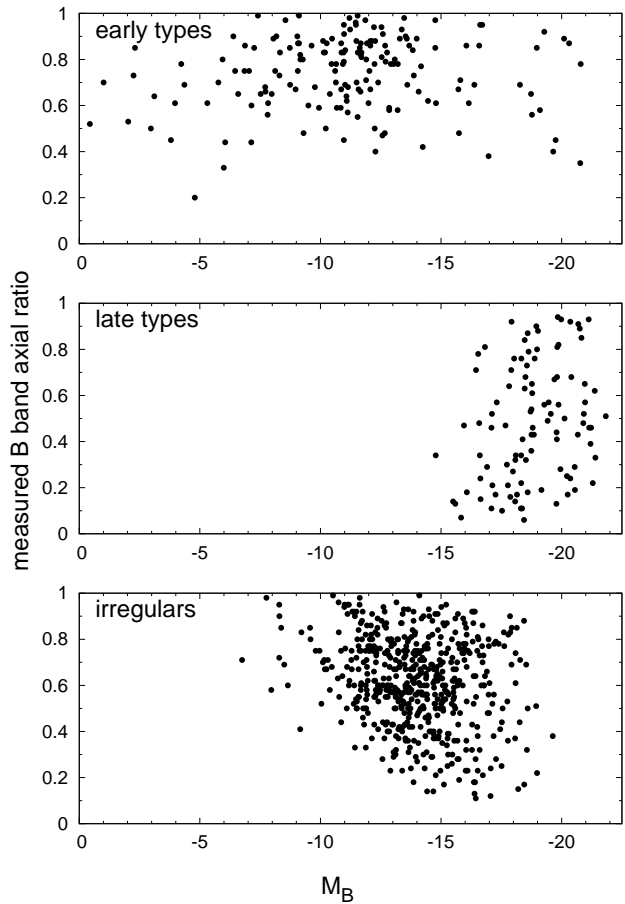,width=3.5truein}
\end{center}
\caption{The B band axial ratios of the three sub-samples of galaxies defined in Table~\ref{tab:groups} plotted against their absolute blue magnitudes.}
\label{fig:all}
\end{figure}

As mentioned above, our sample is drawn from the updated Local Volume catalog, compiled by \citet{kar12}. The catalog consists of all galaxies with radial velocity with respect to centroid of Local Group ${\rm V_{LG}~<~600~km~s^{-1}}$ or distance ${\rm D~<~11~Mpc}$.
Optical axial ratios were measured at the Holmberg isophote (${\rm \sim 26.5^m~arcsec^{-2}}$) in the B-band.
For some dwarf galaxies with extremely low surface brightness (especially the ones which are resolved into individual stars) even the central brightness is fainter than the Holmberg isophote. For these galaxies the axial ratio were measured from the ellipse whose major axis is equal to the scale-length of the brightness profile for that particular galaxy.
Morphological types according to the classification by \citet{deVac91} were used to separate the galaxies into the three primary sub-samples listed in Table~\ref{tab:groups}. We label the three sub-samples as
``early types'' (types $-$8 to 0, inclusive), ``late types''  (types 1 to 7, inclusive) and ``irregulars'' (types 8 to 10, inclusive).  The distributions of the observed axial ratios of the galaxies in the three sub-samples with ${\rm M_B}$ are shown in Figure~\ref{fig:all}.

\begin{table}
\begin{center}
\caption{The sub-samples}
\label{tab:groups}
\begin{tabular}{|lcccc|}
\hline
Sub-Sample&~~~~~~&de Vacouleurs'&${\rm M_B}$&number\\
~&&types&range$^a$&galaxies\\
\hline
\hline
Early types&&$-$8 to 0&&170\\
Late types&&1 to 7&&101\\
Irregulars&bright &8, 9, 10&$-$14.8 to $-$19.6&182\\
Irregulars&intermediate&8, 9, 10&$-$12.6 to $-$14.8&234\\
Irregulars&faint&8, 9, 10&$-$06.7 to $-$12.6&132\\
\hline
\hline
\end{tabular}
\end{center}
\begin{flushleft}
$^a$ for dividing `irregulars' into further sub-samples.
\end{flushleft}
\end{table}

\begin{figure*}
\begin{center}
\psfig{file=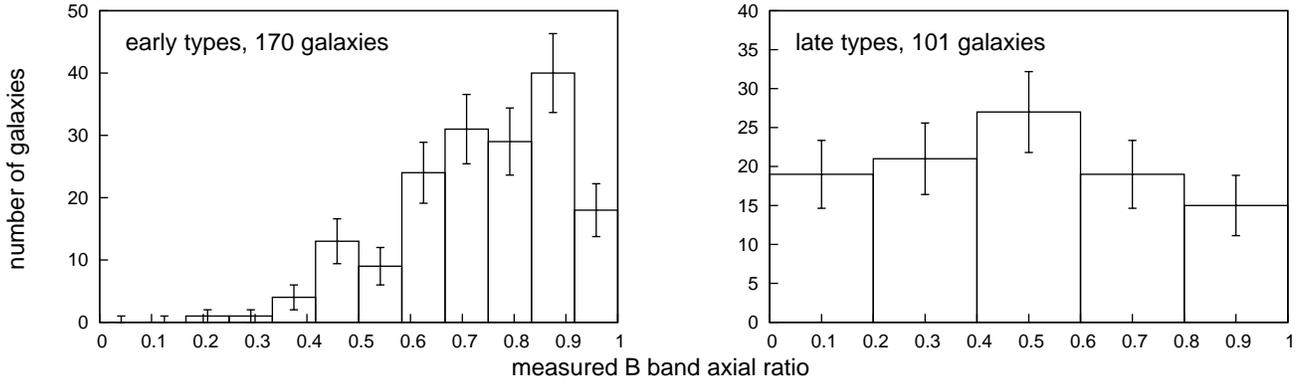,width=7truein}
\end{center}
\caption{Histograms of `early type' and `late type' galaxies binned according to their observed B band axial ratios, with number of bins decided according to the Freedman-Diaconis rule. Error bars shown correspond to ${\rm \sqrt{N}}$~where N is the number of galaxies in the corresponding bin.}
\label{fig:optaxrat_histoth}
\end{figure*}

One can see from Figure~\ref{fig:all} that the faintest galaxies below ${\rm M_B~\sim~-7}$ are all of early type. The distribution of their axial ratios also does not appear to vary much with magnitude, and  has a cut off below an observed axial ratio of $\sim 0.4$. A histogram of the axial ratio distribution is shown in Figure~\ref{fig:optaxrat_histoth} (left panel). The bin widths of the histograms in this figure were determined using the Freedman-Diaconis rule \citep{fre81}. 
This choice of bin width minimizes the mean square error between the observed histogram bar height and the probability density of the underlying distribution.
This observed axial ratio distribution is consistent with what one expects from a population of thick oblate spheroids \citep[e.g. compare with Figures 2(a) and 2(b) from][]{lam92}. 

For late type galaxies, Figure~\ref{fig:all} (middle panel) shows that the observed axial ratio can be quite small ($\lesssim 0.2$). The histogram of the observed axial ratio distribution is shown in the right panel of Figure~\ref{fig:optaxrat_histoth}. The distribution is flat, as expected from a population of thin discs.

\begin{figure}
\begin{center}
\psfig{file=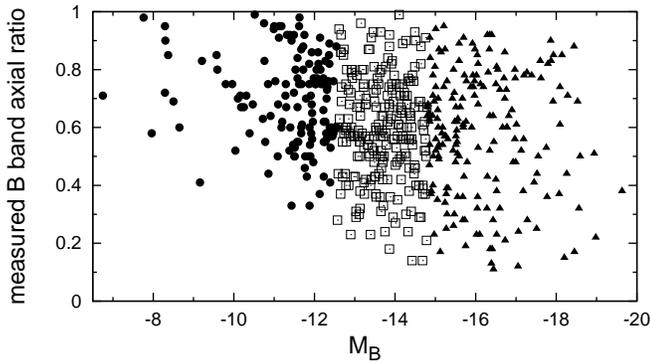,width=3.5truein}
\end{center}
\caption{The B band axial ratios of the 3 sub-samples of  irregular galaxies. The bright irregulars are shown as triangles, the intermediate irregulars as hollow squares and the faint irregulars as filled circles.  See Table~\ref{tab:groups} for the definitions of the sub-samples.}
\label{fig:optaxrat_arir}
\end{figure}

In the rest of the paper we focus on the last sub-sample of galaxies, viz. the `irregular' galaxies. The observed axial ratios of the galaxies in the sub-sample are shown in the bottom panel of Figure~\ref{fig:all}. At the bright end quite small axial ratios are seen, and there appears to be a trend of decreasing minimum observed axial ratio with increasing luminosity. To look at this in more detail, we divide the irregular galaxy sub-sample into three further sub-samples. We clarify that this sub-division is done to understand the variation of shape with intrinsic brightness, although one would expect that the shapes vary smoothly with brightness. These sub-samples consist of the ``bright'' irregulars ($-19.6 < {\rm M_{\rm B}} < -14.8$), the ``intermediate'' irregulars ($-14.8 < {\rm M_{\rm B}} < -12.6$) and the ``faint'' irregulars ($-12.6 < {\rm M_{\rm B}} < -6.7$). There are between 130 to 230 galaxies in each sub-sample (henceforth, by ``sub-samples'' we mean these three sub-samples of irregular galaxies), significantly larger than what has been used in earlier studies.  The observed axial ratio distributions for these three sub-samples are shown in Figure~\ref{fig:optaxrat_arir}.  For the bright sub-sample (which contains many Sdm and Sm type galaxies) very small axial ratios ($\lesssim 0.2$) are found. Below we take a more detailed look at the implications of these observed axial ratio distributions on the intrinsic shapes of the galaxies. 

\section{Methodology}
\label{ssec:method}

We use Monte Carlo simulations to determine the distribution of intrinsic axial ratios which best matches the observed distribution of axial ratios. The procedure we use is the oft-used parametric approach (described below) for determining the intrinsic shapes of non-axisymmetric objects given by \citet{lam92}.
The parameter space describing the distribution of intrinsic axial ratios is gridded and searched, and at each point in the grid a statistically large number of observed axial ratios are simulated and their distribution determined. $\chi^{\rm 2}$ minimization is used to determine which point in the intrinsic axial ratio parameter space grid best fits the observed distribution. 
 
We model the galaxies as triaxial ellipsoids, with `p' and `q' representing the ratios of the middle and minor axis with the major axis of the ellipsoid respectively.  The two axial ratios of any galaxy are assumed to be drawn randomly from two normalized Gaussians of the form:
\begin{equation}
{\frac{1}{\sqrt{2\pi}\sigma_p}}\exp[{\frac{-(p-p_0)^2}{2\sigma_p^2}}]~~\&~~{\frac{1}{\sqrt{2\pi}\sigma_q}}\exp[{\frac{-(q-q_0)^2}{2\sigma_q^2}}]
\label{eqn:optaxrat_i1}
\end{equation}
\noindent
with the condition that 0$<$q$_{\rm 0}$$<$p$_{\rm 0}$$\le$1.
If the randomly chosen p and q using equation~\ref{eqn:optaxrat_i1} are such that p$<$q, then p and q are interchanged.
p$_{\rm 0}$, q$_{\rm 0}$, $\sigma_{\rm p}$ and $\sigma_{\rm q}$~are all varied between 0 and 1.
p$_{\rm 0}$ and q$_{\rm 0}$, being the parameters of interest, are varied in steps of 0.01 whereas $\sigma_{\rm p}$ and $\sigma_{\rm q}$~in steps of 0.05.
For two special cases the triaxial model reduces to spheroidal model, viz. for q$_{\rm 0}$=p$_{\rm 0}$ (prolate) and p$_{\rm 0}$=1 with q$_{\rm 0}<$1 (oblate).

Following \citet{sta77}, if one rotates the triaxial disc along two Euler angles, one can simulate all possible viewing angles.
For each set of p$_{\rm 0}$, q$_{\rm 0}$, $\sigma_{\rm p}$, $\sigma_{\rm q}$, we simulate 10$^{\rm 5}$ galaxies, each with a p and q chosen randomly and as per their assumed distribution, and two viewing angles ($\theta$ varying between 0 to $\pi$, and $\phi$ varying between 0 to 2$\pi$).
The two viewing angles are chosen so as to populate the surface of a unit sphere uniformly using the algorithm proposed by \citet{mar72}.
For each set of p, q and two viewing angles ($\theta$ and $\phi$), the apparent axial ratio $\beta$ is calculated using formulae given in \citet{sta77} and \citet{bin80} which are as follows:
\begin{equation}
\beta~=~\sqrt{\frac{j+l-\sqrt{(j-l)^2~+~4k^2}}{j+l+\sqrt{(j-l)^2~+~4k^2}}}~,
\label{eqn:optaxrat_i2}
\end{equation}
\noindent
where
\begin{equation}
\begin{array}{cr}
j~=~q^2 sin^2 \theta~+~p^2 sin^2 \phi~ cos^2 \theta~+~cos^2 \phi~ cos^2 \theta~,\\
k~=~(1~-~p^2)sin \phi~ cos \phi~ cos \theta~,\\
l~=~p^2 cos^2 \phi~+~sin^2 \phi~.
\end{array}
\label{eqn:optaxrat_i3}
\end{equation}
The 10$^{\rm 5}$~simulated apparent axial ratios are then binned into number of bins equal to the that used to create the histogram of observed axial ratios for the particular sub-sample of galaxies in question, to obtain an apparent axial ratio distribution for each set of p$_{\rm 0}$, q$_{\rm 0}$, $\sigma_{\rm p}$, $\sigma_{\rm q}$.
Finally, each such binned simulated distribution is compared to the binned observed distribution for that particular sub-sample, and the distribution which gives the minimum $\chi^{\rm 2}$~is identified.

The following method was adopted in order to determine the number of bins to be used for each sub-sample when binning the observed axial ratios.
The initial bin widths were determined according to the Freedman-Diaconis rule \citep{fre81}.
After following the entire method discussed above, if the number in each bin of the simulated distribution with the minimum $\chi^{\rm 2}$~did not match the number in the same bin of the observed distribution within the Poisson errors due to binning of the data, the number of bins was increased by one and the method repeated.
In this way, we found that for all the sub-samples binning the observed axial ratios in 8 bins was optimal.

\section{Results and Discussion}
\label{sec:optaxrat_res}

\begin{figure}
\begin{center}
\psfig{file=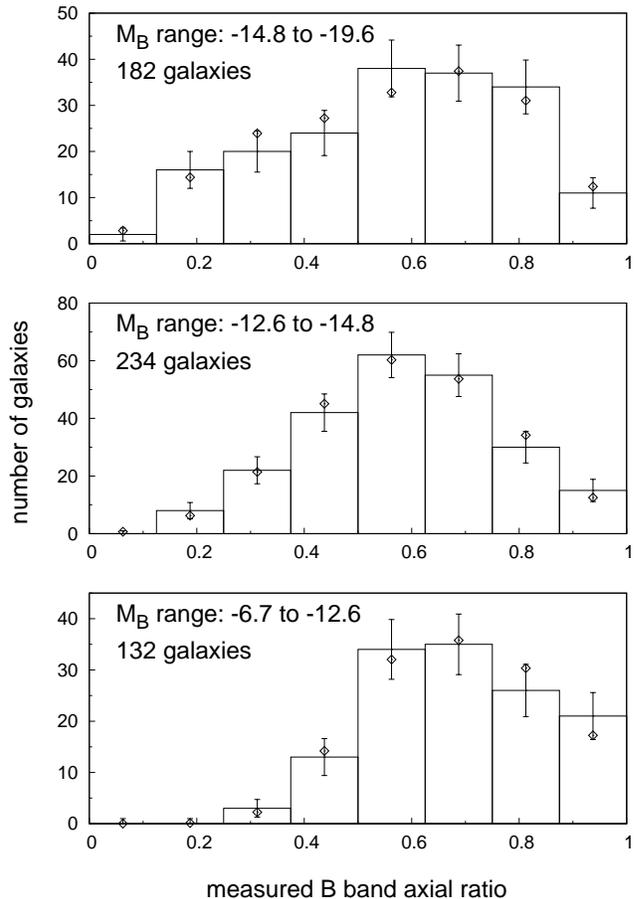,width=3.5truein}
\end{center}
\caption{Histograms of the 3 sub-samples of irregular galaxies binned according to their observed B band axial ratios. Error bars shown correspond to ${\rm \sqrt{N}}$~where N is the number of galaxies in the corresponding bin. The open diamonds represent the best fit to the apparent axial ratio data obtained from the simulations.}
\label{fig:optaxrat_histir}
\end{figure}

\begin{figure*}
\begin{center}
\psfig{file=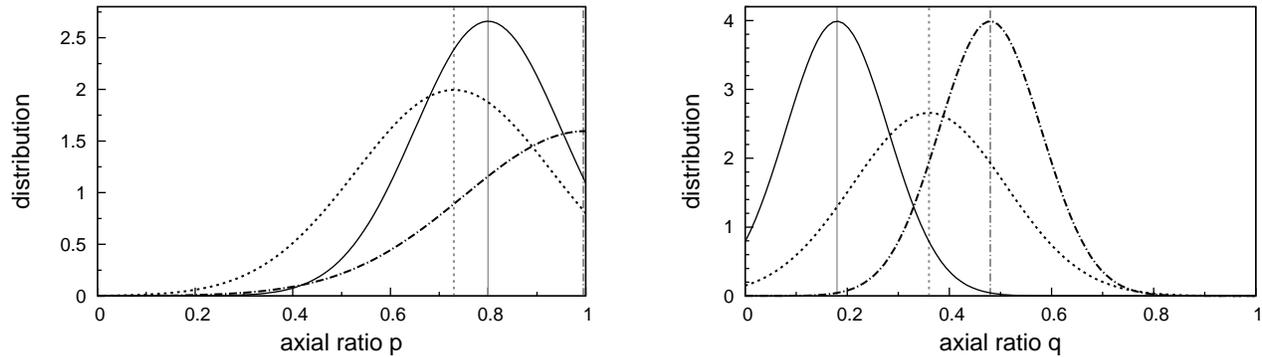,width=7truein}
\end{center}
\caption{The distributions of intrinsic axial ratios for the best fits shown in Figure~\ref{fig:optaxrat_histir} for the 3 sub-samples of irregular galaxies. The Gaussian distributions for the two axial ratios p and q are shown separately. In each panel, the bold line is for the sub-sample of brighter irregulars, the dotted line is for the sub-sample of intermediate luminosity irregulars, and the dot-dashed line is for the sub-sample of fainter irregulars. The vertical grey bold, dotted and dot-dashed lines mark the positions of the peak of the Gaussians corresponding to the brighter, intermediate luminosity and fainter irregular sub-samples respectively.}
\label{fig:optaxrat_inar}
\end{figure*}

\begin{figure}
\begin{center}
\psfig{file=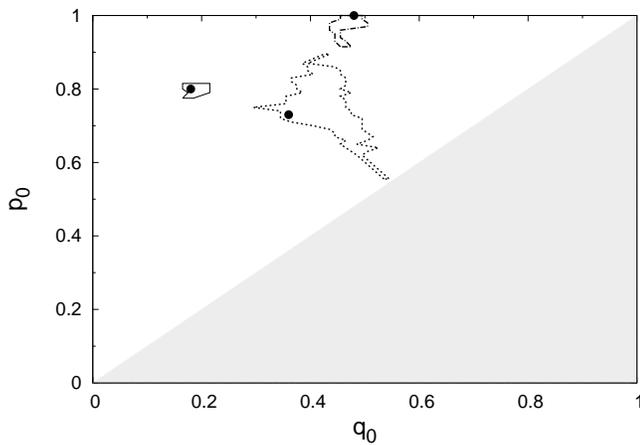,width=3.5truein}
\end{center}
\caption{The best fit p$_{\rm 0}$~and q$_{\rm 0}$, the two parameters of interest, are shown for the three subsamples (filled circles) along with the corresponding confidence intervals: bold line for the brighter, dotted line for the intermediate luminosity and the dot-dashed line for the fainter irregulars respectively.  The faint grey area represents the p$_{\rm 0}$-q$_{\rm 0}$ parameter space excluded from the search, as it is equivalent to flipping p$_{\rm 0}$ and q$_{\rm 0}$.}
\label{fig:ci}
\end{figure}

The best fit simulated distributions to the observed distributions
for each of the sub-samples are shown in Figure~\ref{fig:optaxrat_histir}. The intrinsic distributions corresponding to these apparent distributions are given in Table~\ref{tab:vals}, and are shown in Figure~\ref{fig:optaxrat_inar}.
For estimating the error on the parameters, for each of the three sub-samples we define the confidence interval as the parameter space around the best fit parameters which give rise to simulated distributions that match the observed binned distribution at each point within Poisson errors due to binning. 
The extent of each parameter within confidence intervals thus defined for each of the three sub-samples are given in Table~\ref{tab:vals}.
The confidence intervals for p$_{\rm 0}$~and q$_{\rm 0}$, the two parameters of interest are also shown in Figure~\ref{fig:ci}. The confidence intervals suggest a continuous variation of the two intrinsic axial ratios with changing luminosity, and our choice of sub-samples has only discretized this variation.

\begin{table}
\begin{center}
\caption{Results}
\label{tab:vals}
\begin{tabular}{|ccccccc|}
\hline
Group&p$_{\rm 0}$&$\sigma_{\rm p}$&q$_{\rm 0}$&$\sigma_{\rm q}$\\
${\rm M_B}$ range&&&&&\\
\hline
\hline
$-$14.8 to $-$19.6&0.80$^{+0.02}_{-0.03}$&0.15$^{+0.03}_{-0.03}$&0.18$^{+0.04}_{-0.02}$&0.10$^{+0.08}_{-0.03}$\\
$-$12.6 to $-$14.8&0.73$^{+0.2}_{-0.2}$&0.20$^{+0.2}_{-0.08}$&0.36$^{+0.2}_{-0.07}$&0.15$^{+0.1}_{-0.08}$\\
$-$06.7 to $-$12.6&1.00$^{}_{-0.09}$&0.25$^{+0.03}_{-0.08}$&0.48$^{+0.03}_{-0.05}$&0.10$^{+0.03}_{-0.03}$\\
\hline
\hline
\end{tabular}
\end{center}
\end{table}

This analysis quantifies what was discernible from the overall trends, viz. that irregular galaxies become thicker with fainter luminosities.
The conclusions one can draw from  Figure~\ref{fig:optaxrat_inar} are:\\
(i) The bright irregular galaxies have slightly elliptical discs. That is they have triaxial shapes, with the two larger axes having slightly differing lengths, and the ratio of the smallest axis to the two larger axis being small and comparable to that for normal spirals.\\
(ii) The intermediate irregulars have somewhat thicker elliptical discs. Their larger and smaller intrinsic axial ratios peak at $\sim$0.7, and $\sim$0.4 respectively.\\
(iii) In contrast to their brighter counterparts, the faint irregular galaxies are thick oblate spheroids, with the intrinsic axial ratio being $\sim$0.5. Interestingly, their neutral hydrogen discs also appear to be puffy oblate spheroids with similar axial ratio \citep{roy10}. 

The increasing thickness of dwarf galaxy discs with decreasing luminosity can be understood in terms of the increasing importance of turbulent pressure support in stabilizing these galaxies. In the faintest systems, the random motions can be comparable to the rotational speeds \citep[see e.g.][]{beg04b,beg06}. Interestingly this trend towards thicker stellar discs has been reproduced in recent simulations of dwarf galaxy formation which include feedback due to star formation \citep[see e.g.][]{kau07,gov10}. 
One could conjecture that the tri-axiality of the brighter irregular
galaxies reflects the fact that their star formation is patchy, leading
to a high likelihood of the galaxy not having azimuthal symmetry.
Interestingly however, \citet{lam92} found that the discs of brighter spiral galaxies have ellipticity similar to what we find here for the irregular galaxies. The ellipticity of spiral discs  has been confirmed by recent SDSS based
studies of spiral galaxies \citep{ryd04,pad08}.
\citet{lam92} suggest that a triaxial potential (due to dark matter) in the halos of large spirals as one of the possible causes for the difference in the scale-lengths of the two larger axes. Given the general continuity in
galaxy properties, one would expect that the same mechanism could be
responsible for the ellipticity that we observe in our sample. 

In summary we use the compilation of galaxy axial ratios available in the Local Volume catalog, to determine the intrinsic shape of late type galaxies (viz. de Vaucouleurs types 8, 9 and 10). The brightest
of these galaxies have thin discs, and the discs get thicker as one
moves to fainter galaxies.

\bsp

\label{lastpage}

\end{document}